\pdfoutput=1
\documentclass[11pt, a4paper]{article}
\usepackage{jcappub.local}

\usepackage{subfigure}
\usepackage[subfigure]{graphfig}

\usepackage[section]{placeins}
\makeatletter
\AtBeginDocument{%
  \expandafter\renewcommand\expandafter\subsection\expandafter{%
    \expandafter\@fb@secFB\subsection
  }%
}
\makeatother

\title{Non-Uniqueness of Classical Inflationary Trajectories on a High-Dimensional Landscape}

\author[a]{Junyu Liu}
\author[*b]{Yi Wang}
\author[b]{Siyi Zhou}
\affiliation[a]{University of Science and Technology of China, \\
Hefei, Anhui 230026, P.R.China}
\affiliation[b]{Department of Physics, The Hong Kong University of Science and Technology,\\
Clear Water Bay, Kowloon, Hong Kong, P.R.China}
\emailAdd{phyw@ust.hk}

\abstract{Motivated by the string landscape, the inflationary potential may be high dimensional and complicated. We propose a new way to construct some high dimensional random potentials, and simulate inflationary dynamics on top of that, for up to 50-dimensions in field space. Especially, random bifurcations of classical inflationary trajectory are studied. It is shown that the bifurcation probability increases as a function of number of dimensions. Those random bifurcations are not consistent with observations, and dramatically limit the parameter space of inflation on a complicated landscape. For example, in 10 dimensions, only $10^{-3} \sim 10^{-6}$  of the parameter space volume leads to unique classical trajectories. The rest is ruled out by random bifurcations.}

\begin{document}

\maketitle

\section{Introduction} \label{sec:intro}
Inflation is the leading paradigm for the very early universe cosmology. However, the detailed dynamics of inflation still remains mysterious. We have not confirmed whether the inflationary potential is single or multiple dimensional, smooth or bumpy, featuring straight trajectory or curved, etc.

The situation is rapidly improving with the on-going and future experiments. With more data on the cosmic microwave background (CMB) and the large scale structure (LSS), we expect to have better confidence in addressing the above questions. On the other hand, it is also important to better understand the underlying theoretical construction of inflation, so that the naturalness of those models can be better addressed.

One of the particularly interesting ideas for theoretically constructing inflationary models is the string landscape. The string landscape predicts that string theory has very complicated vacuum structure. A large number of light fields can be present in the string landscape, unless stabilized explicitly. The number of those fields can easily reach $\mathcal{O}(100)$. Inflation on such a random potential has been studied in \cite{Berera:1996nv,Huang:2008jr, Tye:2009ff, Battefeld:2011yj, Dias:2012nf, McAllister:2012am, Gao:2014uha, Green:2014xqa}.

To study inflation on a random potential, the first step would be to construct such a random potential numerically \footnote{There are analytical studies of statistical features of such random potentials. Even in those case, it would be helpful to numerically test the analytical calculation, and extend it where perturbative methods break down, with numerical methods.}. There have been a few methods for random potential generation in the literature. One can either generate random numbers in position space and interpolate \cite{Duplessis:2012nb}; or generate Fourier coefficients and then Fourier transform to field space \cite{Tegmark:2004qd, Frazer:2011br}; or making use of Dyson Brownian motion \cite{Marsh:2013qca, Battefeld:2014qoa}; or use Gaussian covariance function to generate the potential dynamically \cite{Bachlechner:2014rqa}.Those methods, looking from a different angle, can also be classified into two types:

\begin{itemize}
\item Static methods: A whole patch of inflationary potential is generated (or at least determined \footnote{It may be interesting to investigate the possibility to determine, instead of actually generate the whole potential. The time and space complexities may be greatly reduced. But such algorithms are not available, to the best of our knowledge.}) before evolving the equation of motion of the inflaton. The pros and cons of static methods of potential generation include
\begin{itemize}
\item The potential is non-perturbative in the sense that we are not Taylor expanding the potential around any point.
\item There is no bias from different classical trajectories of inflation, because the potential is trajectory-independent.
\item Scenarios such as bifurcation, or self-crossing of inflationary trajectory can be studied.
\item Unfortunately, it is computationally extremely expansive to generate such a potential in high dimensional field space. Because the computational complexity to generate such a potential is $\mathcal{O}(e^N)$. To calculate such a fully random potential, the exponential computational expense is unavoidable because even to generate random numbers on the vertices of an $N$ dimensional cube needs $2^N$ random numbers. Interpolation of those numbers to obtain continuous fields is another $\mathcal{O}(e^N)$ calculation.
\end{itemize}

\item Dynamical methods: The potential is dynamically generated along an inflationary trajectory. The advantages and disadvantages of dynamical methods are to the opposite compared with static ones:
\begin{itemize}
\item Fast computation ($\mathcal{O}(N^2)$ for the existence of second derivative).
\item The potential only applies for a neighbourhood of the inflationary trajectory, because the dynamical methods depend on Taylor expanding around the inflationary trajectory.
\item It is hard to study scenarios such as bifurcation, or self-crossing of inflationary trajectory. For example, if the trajectory is self-crossing, the potential at the self-crossing point shall have different values and derivatives when the inflaton runs into this point from different trajectories. On the other hand, it is interesting to note that after taking special care for the consistency of crossing points, it is possible to study such behaviors using dynamical methods \cite{Bachlechner:2014rqa}.
\end{itemize}

\end{itemize}

In the present paper, we investigate bifurcations in a high-dimensional random potential. In section \ref{sec:gener-rand-potent}, we propose a new approximation method of potential generation, combining the advantages of static and dynamical methods.

\begin{figure}[htbp]
  \centering
  \includegraphics[width=0.7\textwidth]{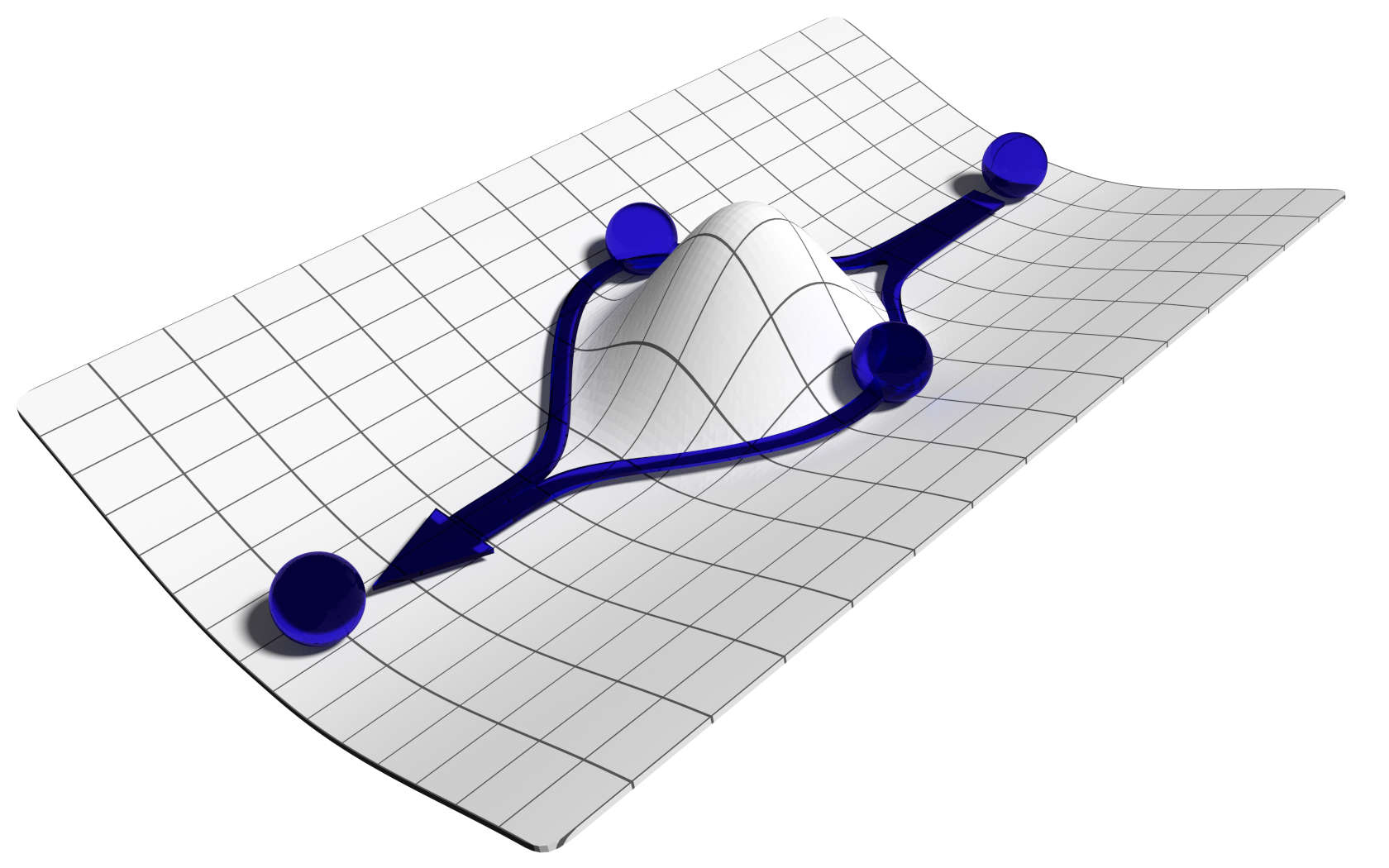}
  \caption{\label{fig:bifurcation} Multi-stream inflation with bifurcations of classical inflationary trajectory.}
\end{figure}

After the construction of the random potential, we move on to study bifurcations. When hitting bumps in the potential, the inflationary trajectory may bifurcate. The situation is illustrated in Fig.~\ref{fig:bifurcation}. The situation is known as multi-stream inflation \cite{Li:2009sp}.

Bifurcations of inflationary trajectory, if under control (for example, follows from spontaneous symmetry breaking), can provide explanations of CMB statistical asymmetries \cite{Wang:2013vxa} and/or the CMB cold spot \cite{Afshordi:2010wn}. This is because the e-folding number difference between different patches of the universe are slightly different. Moreover, temporary domain walls form between different trajectories before those trajectories recombine, which contributes to CMB statistical anisotropies \cite{Jazayeri:2014nya}. The dynamics of eternal inflation can also be affected by such bifurcations \cite{Li:2009me, Wang:2010rs}.

However, if the bifurcations happens during observable inflation, while not under control, as in the case here on a random potential, disasters happen \cite{Duplessis:2012nb}. This is because random bifurcations typically have e-folding number difference of order 1 in different branches. Thus the CMB, observed at different directions, shall look highly different from each other, with $\Delta T /T \sim \mathcal{O}(1)$. This contradicts the CMB/LSS observations on large scales, and primordial black holes form after such scales return to the horizon, making contradicting predictions for small scale physics all the way down to reheating.

\begin{figure}[htbp]
  \centering
  \includegraphics[width=0.7\textwidth]{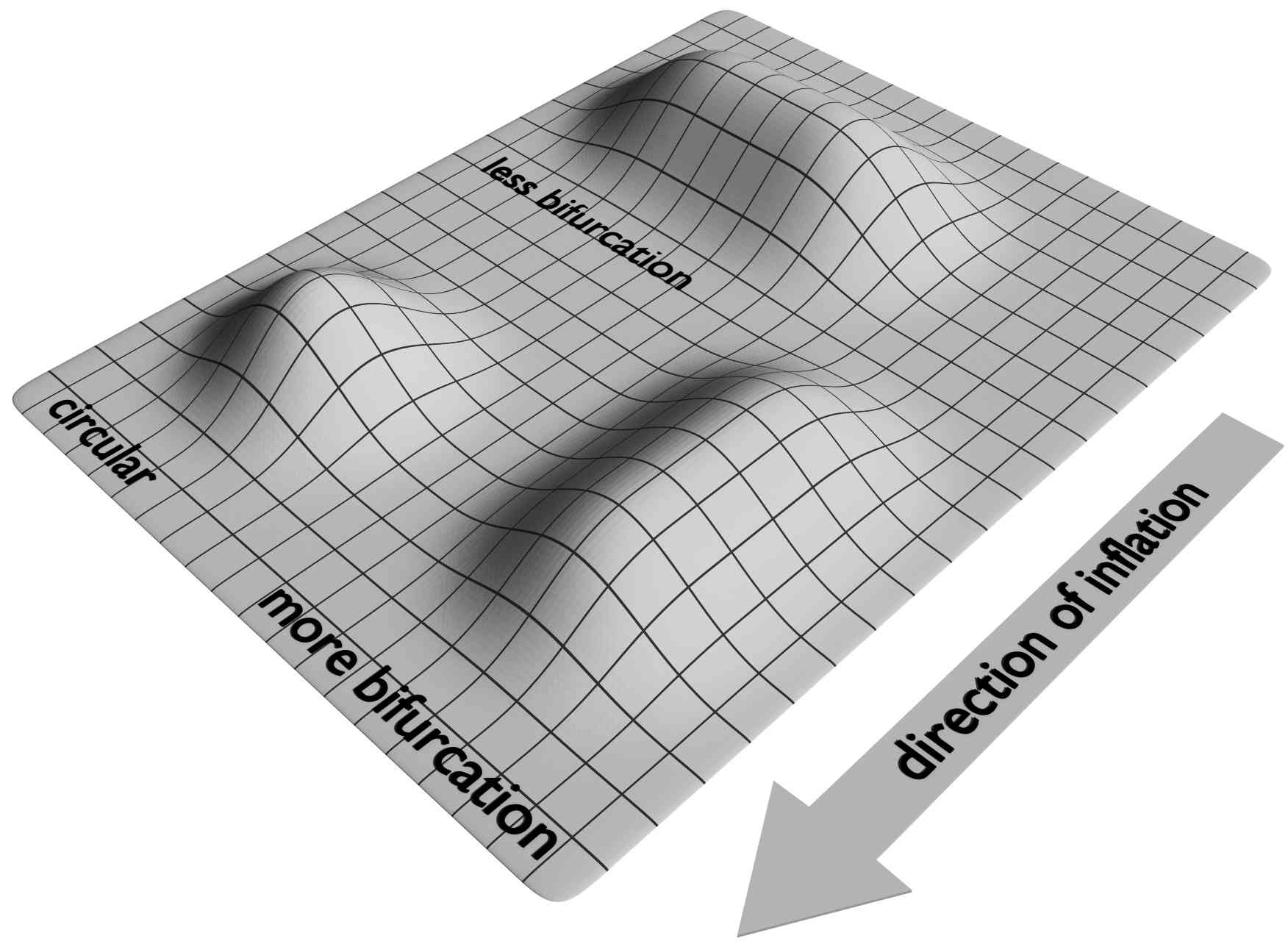}
  \caption{\label{fig:ellipticity} Relation between ellipticity and bifurcation probability.}
\end{figure}

Thus we should not allow random bifurcations to happen. As a result, inflation on a random potential is constrained. In \cite{Duplessis:2012nb}, the bifurcation constraint on a two dimensional potential is considered. We find that the constraint is pretty weak for the case of a two dimensional potential, if the characteristic lengths of the random features on the inflation and isocurvature directions are equal (i.e. the random features are statistically circular shaped). However, if there is an ellipticity, such that the random feature is statistically more than 10 times longer in the inflation direction, bifurcations can happen easily. The situation is illustrated in Fig.~\ref{fig:ellipticity}. Examples of such potential with ellipticity can be found in \cite{Long:2014dta}.

In this paper, we shall keep the features statistically circular, and consider higher dimensional random potentials. The method of calculation is similar to that of \cite{Duplessis:2012nb}, which is reviewed in Section \ref{sec:dynamics-inflation}. In Section \ref{sec:bifurc-rand-potent}, we present the result of bifurcation constraint up to 50 field dimensions, for large and small field inflation models respectively. We conclude in Section \ref{sec:conclusion-outlook}.
\section{Generation of random potential} \label{sec:gener-rand-potent}
In this section, we propose a method for high-dimensional random potential generation. Our purpose is to combine the advantages of static and dynamical methods, and enable study of bifurcation on a high dimensional random potential.

We shall use
\begin{align}
  V = V_0(\phi_1) + \sum_{i < j} \delta V_{i,j}(\phi_i, \phi_j)~,
\end{align}
where $\phi_1$ is the inflation direction with a slope, and the other directions are flat up to random noises $\delta V_{i,j}(\phi_i, \phi_j)$.

If we were to use such a multiple dimensional potential for a general purpose, there shall be serve draw-backs, because the potential does not have enough ``randomness''. For example, consider 3-dimensional case. We have
\begin{align}
  V = V_0(\phi_1) + \delta V_{1,2}(\phi_1, \phi_2) + \delta V_{1, 3}(\phi_1, \phi_3) + \delta V_{2, 3}(\phi_2, \phi_3)~.
\end{align}
We observe that $\partial_{\phi_1}\partial_{\phi_2}\partial_{\phi_3}V = 0$, which is not true for a fully random potential. Also, $\partial_{\phi_2}\partial_{\phi_3}V$ depends on $\phi_1$, which is also unpleasant.

However, for the case of inflation with one dominant direction, such draw-backs shall not cause problems for practical purposes. Because for the purpose of studying bifurcations, we shall only be interested in the first derivative of the potential. (And the potential can easily be generated into a summation of $V_{i,j,k}(\phi_i, \phi_j, \phi_k)$ or higher orders if needed.)

Although the current method is a simplified version of static potential generation method, it is also worth mentioning the connection with the dynamical potential generate methods. During inflation, the inflaton field rolls much faster compared with other massless or stabilized directions \footnote{If there are a few classically rolling directions, one can always rotate the fields such that only one field rolls classically. Moreover, a turning of trajectory can be parametrized as interactions between different directions with straight trajectory from non-trivial field space metric \cite{Chen:2009we, Chen:2009zp}.}. This is because, at each e-fold, the inflaton rolls a distance of $\dot\phi / H$, plus small quantum fluctuations. On the other hand, the other directions moves a distance of $H/(2\pi)$ because of quantum fluctuations. The ratio of those amplitudes is $1/\sqrt{P_\zeta} \sim 10^{5}$.

As a result, effectively, $\phi_1$ acts as the time variable along the inflationary trajectory. Thus the method can be considered effectively as dynamical potential generation, where $\phi_1$ parametrizes the inflationary trajectory. However, thanks to the static nature, the current method shall not suffer from the limitations of local Taylor expansion, and can be used to study bifurcation events.

The time complexity for this potential generation method is $\mathcal{O}(N^2)$ if only random first derivative is needed (as in our case in later sections). It is worth to note that, the algorithm may further be optimized into $\mathcal{O}(N)$, by noting the above fact that $\phi_1$ parametrizes the inflationary trajectory. One can construct $V = V_0(\phi_1) + \sum_{i} \delta V_i(\phi_1, \phi_i)$. However, if this simplification is made, statistical symmetry of field rotation in $\delta V$ is broken. As a result, if one still hope the random part of the potential to be symmetric, additional tunings are needed. To avoid possible issues, we shall not use this simplification in this paper.

For doing numerical work, we yet need to specify a method to generate $\delta V_{i, j}(\phi_i, \phi_j)$. As mentioned in the introduction, there are many ways to do it. Here we shall first generate $\delta V_{i, j}(\phi_i, \phi_j)$ on a grid, then interpolate to get smooth fields.

\section{Dynamics of inflation} \label{sec:dynamics-inflation}
In this section, we review the inflationary dynamics. On a multi-dimensional landscape, the classical equation of motion for the fields is given by
\begin{align}
  \ddot{\phi_i}+3H\dot{\phi_i}+\partial_iV=0~,
\end{align}
where the potential $V$ is constructed in the previous section. The shorthand notation $\partial_i$ is used to denote the partial derivative with respect to the field $\phi_i$. To simplify the calculation, $H \simeq \sqrt{V/(3M_p^2)}$ is still a good approximation. This is because, on a random potential, in order for the inflationary trajectory not to get stuck, the kinetic energy of the inflaton is not likely to exceed twice the slow roll value, thus still a few percents.

To encode quantum fluctuations, we use Starobinsky's stochastic method \cite{Starobinsky:1986fx}, where quantum fluctuations are considered effectively as source terms for the classical equation of motion. Such source terms can be derived by integrating out the sub-Hubble modes. The quantum-corrected equation of motion reads
\begin{align}
  \ddot{\phi_i}+3H\dot{\phi_i}+\partial_iV = \frac{3}{2\pi}H^\frac{5}{2}\eta_i(t)~,
\end{align}
where the $\eta_i$ term is the random source to denote the quantum corrections and follows independent Gaussian distribution \footnote{For the isocurvature fields, the $\dot\phi_i$ could be small, or even zero without the presence of the RHS. In such a special case, the RHS is the major driving force for this differential equation. (But note that this can never happen in the inflaton direction.) However, this fact does not indicate that the back-reactions are out of control. This is because, the scalar field fluctuation amplitude $H / (2\pi)$ originates from de Sitter expansion. Thus as long as $H$ does not receive large back-reaction (which is always true in our work), the back-reaction to Eq. (3.2) is small and the random source is highly Gaussian.}. The $\eta_i$ terms are normalized as
\begin{align}
  \langle \eta_i(t)\eta_j(t')\rangle=\delta(t-t')\delta_{ij}~,
\end{align}
such that during a Hubble time and averaged on a Hubble volume, the amplitude of quantum fluctuation on each field direction is $H/(2\pi)$.

In the numerical calculation, we have to discretize the time interval as $\Delta t=t_n-t_{n-1}$. While replacing differential equations into difference equations, the Dirac delta function is replaced by
\begin{align}
  \delta(t_a-t_b)\to\frac{\delta_{ab}}{t_{a+1} - t_a} ~,
\end{align}
where $t_n$ denotes the time of the $n$th step.

In the following section, inflation on a random potential is studied based on the above setups.
\section{Bifurcations on a random potential}\label{sec:bifurc-rand-potent}
In this section, we investigate the bifurcation probabilities with different field-space dimensions. As we shall show, the bifurcation probabilities increase as a function of field-space dimension, in both large field and small field inflation models.

\subsection{Large field inflation}

First, we consider large field inflation, with potential $V_0(\phi_1)=\frac{1}{2}m^2{\phi_1}^2$, plus a random source introduced following the algorithm described in Sec. \ref{sec:gener-rand-potent}. The mass parameter is chosen to be $m=6.5\times10^{-6}M_p$. Note that we have picked the $\phi_1$ direction to be the direction with a non-random slope.

Parameters are introduced to parametrize the random part of the potential:
\begin{itemize}
\item We use $\Delta_p \phi_i$ ($i=1,\ldots, n$) to denote the characteristic distance in the field space, between random bumps in the potential. Operationally, we are generating the random potential by interpolation on grids. Thus $\Delta_p \phi_i$ becomes the grid size in the $i$-th direction.

In the two dimensional example \cite{Duplessis:2012nb}, it is shown that the bifurcation probability increases significantly as a function of $\Delta_p \phi_1 / \Delta_p \phi_2$. This is because, when $\Delta_p \phi_1$ increases, bumps in the $\phi_1$ direction are more distant from each other. Thus it is easier to keep the inflationary trajectory slowly rolling. On the other hand, when $\Delta_p \phi_2$ decreases, bumps in the $\phi_2$ direction are narrower. Thus it is easier for a small quantum fluctuation in the $\phi_2$ direction to bifurcate the inflationary trajectory. We expect that this intuition still hold in general $n$-dimensions, simply by replacing $\Delta_p \phi_2$ into $\Delta_p \phi_j$ ($j=2,...n$).

In the numerical study, to reduce the number of parameters, we shall choose $\Delta_p \phi_1 = \Delta_p \phi_2 = \ldots = \Delta_p \phi_n \equiv \Delta$.

\item We use $A_i$ ($i=1, \ldots, n$) to denote the relative amplitude between the random part of the potential and the slope $V_0$, contributed from the $i$-th dimension. For simplicity, we calculate only the case $A_1 = \ldots = A_n \equiv A$. As $A$ increases, the parameter space are parted into three regions:
\begin{itemize}
\item Small $A$: If there is no randomness ($A=0$), we essentially have single field inflation, plus decoupled isocurvature fluctuations. Thus it is natural to expect that for sufficiently small $A$, no bifurcation or stuck behavior takes place.
\item Intermediate $A$: As we shall show, a bifurcation region emerges as $A$ increases. This region is largely overlooked in the previous study of multi-field inflation (except \cite{Duplessis:2012nb}), and should be excluded, because bifurcation introduces too large density fluctuations.
\item Large $A$: When $A$ further increases, the slope $\partial_1 V$ shall not be monotone. Thus the inflationary trajectory gets easily stuck at eternal inflation. We shall exclude this case from our study.
\end{itemize}
\end{itemize}

It is intuitive to think about bifurcations from Fig.~\ref{fig:bifurcation}. Nevertheless, it is important to define the bifurcation events more precisely. We define bifurcation as that, at least one of direction $\phi_j$ ($j=1,...n$) satisfies the condition $ |\phi_j^I-\phi_j^{II}|>\Delta$ at the end of inflation, where $\phi_j^I$ and $\phi_j^{II}$ are two sample trajectories. Two further comments are in order for the definition of bifurcation probability
\begin{itemize}
\item We have excluded the eternal inflation samples when defining the bifurcation probability.
\item The above definition of bifurcation stops making sense when $\Delta \lesssim \sqrt{N_e}H$, where $N_e\sim 60$ is the e-folding number of observable inflation. This is because when $\Delta \lesssim \sqrt{N_e}H$, trivial random fluctuations may be identified as an bifurcation event following the above definition. However, the parameter space that we have considered, in both large field and small field examples, $\Delta$ is orders of magnitude greater than $\sqrt{N_e}H$. Thus this subtlety does not affect our study.
\end{itemize}

\begin{figure}[htbp]

  \includegraphics[width=0.48\textwidth]{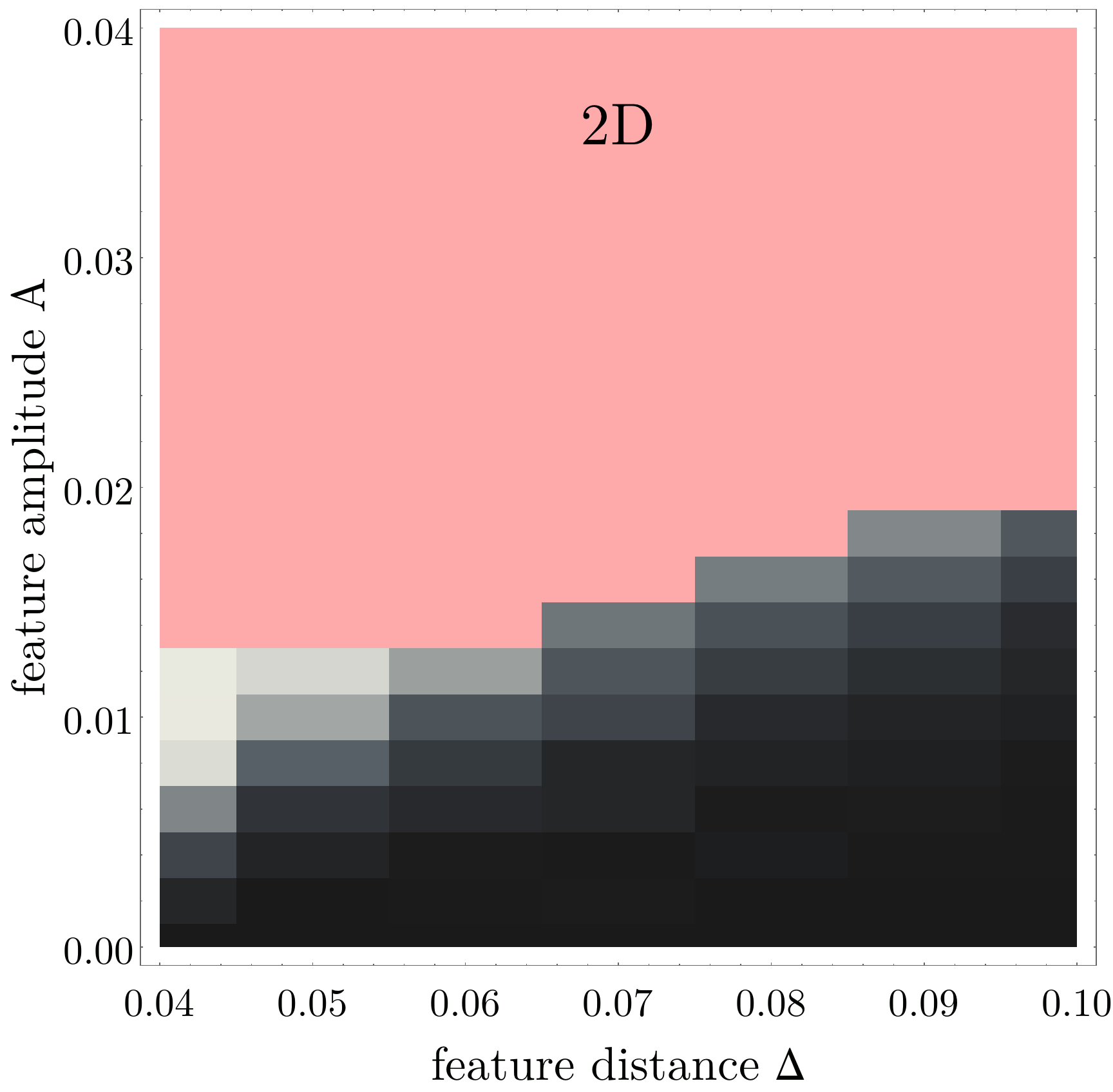}\hspace{0.04\textwidth}
  \includegraphics[width=0.48\textwidth]{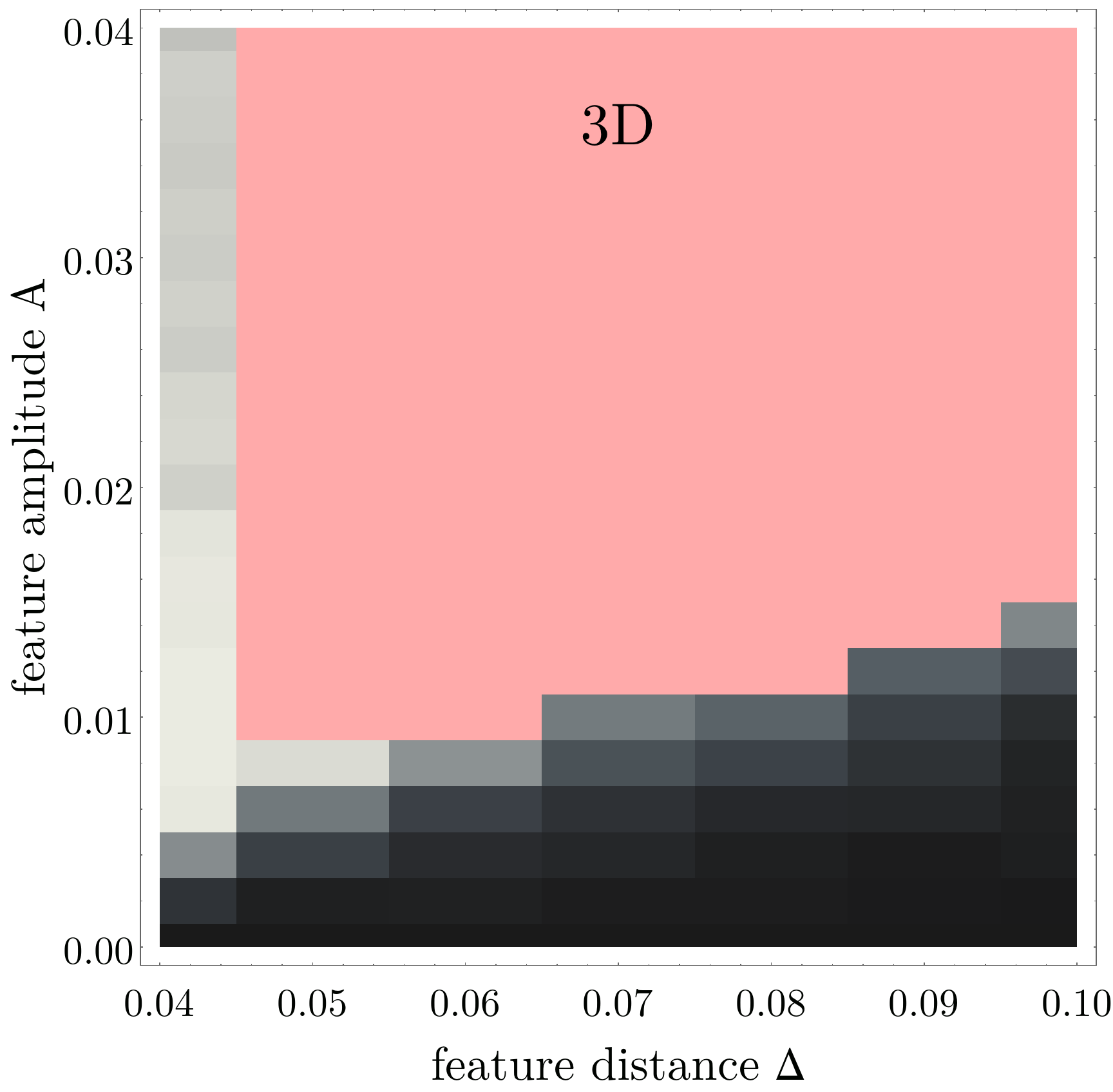}\vspace{0.04\textwidth}

  \includegraphics[width=0.48\textwidth]{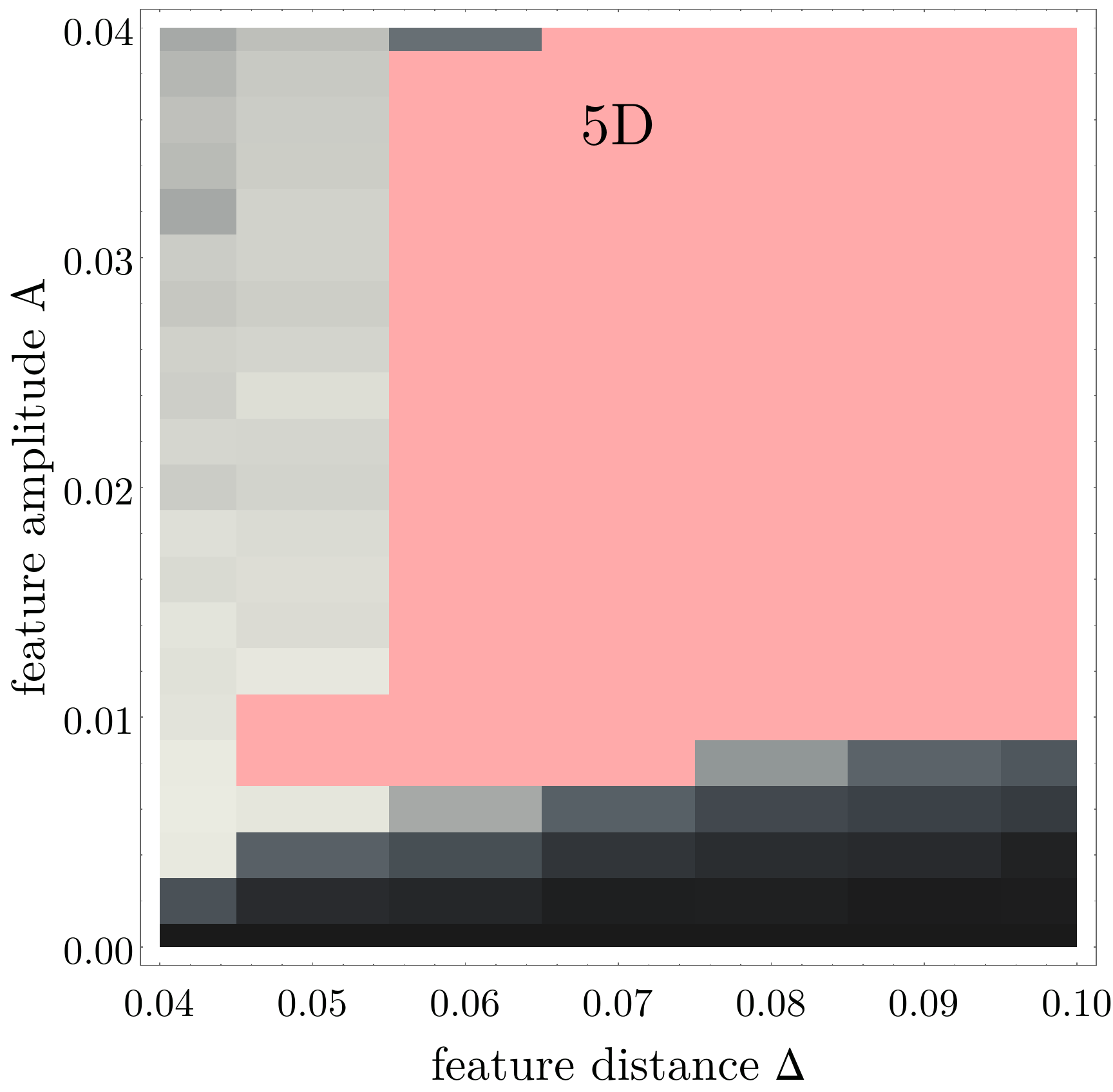}\hspace{0.04\textwidth}
  \includegraphics[width=0.48\textwidth]{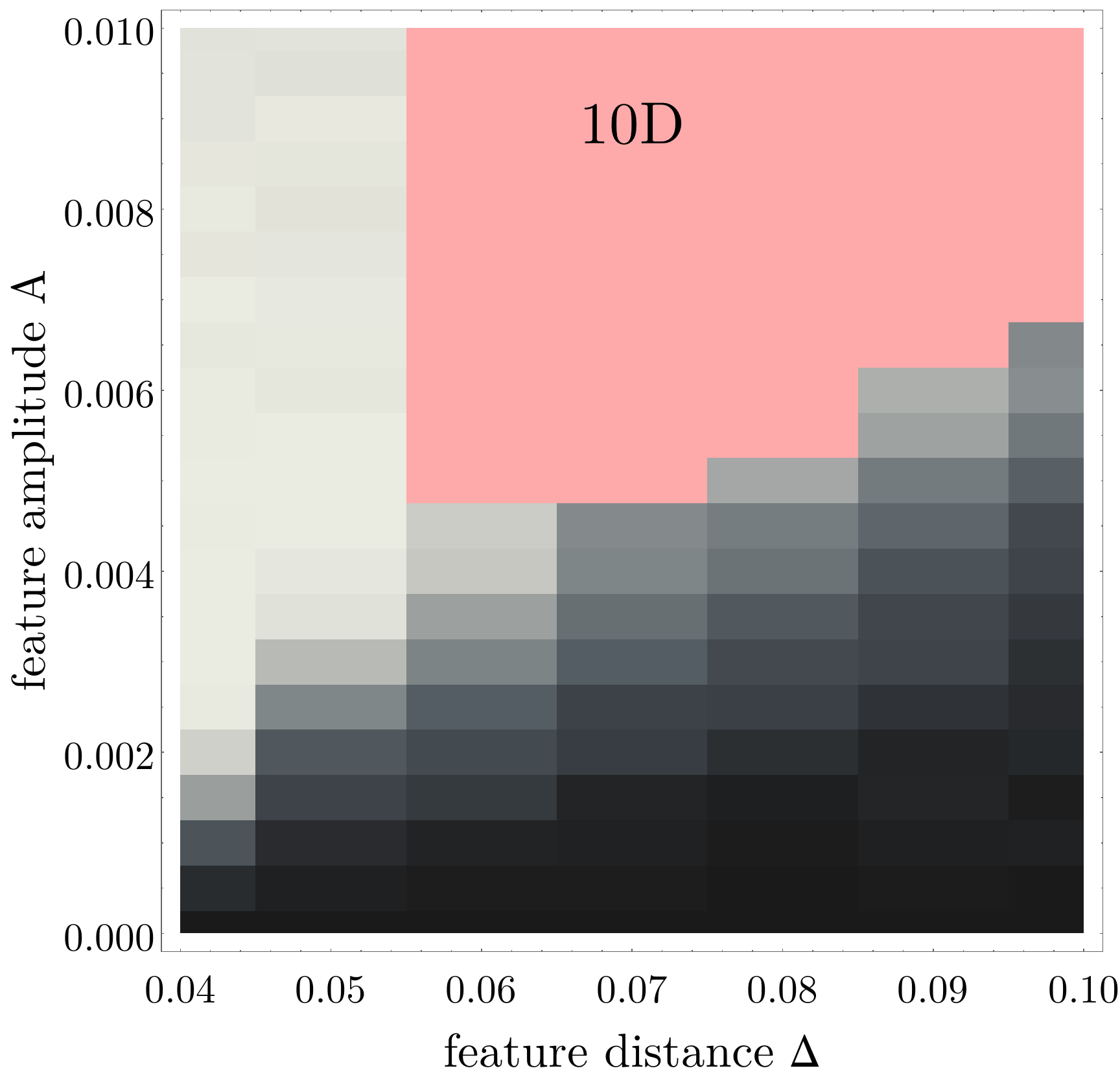}\vspace{0.04\textwidth}

  \hspace{0.25\textwidth}\includegraphics[width=0.5\textwidth]{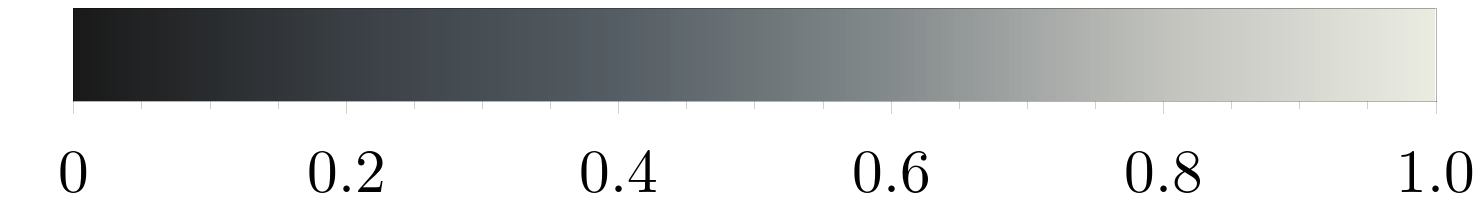}

  \caption{Density plot for bifurcation probabilities in 2, 3, 5 and 10 dimensions. The pink regions corresponds to exclusion regions where the inflationary trajectory is stuck at eternal inflation.
Notice that in the 10-dimensional case, as the stuck region gets bigger, we have zoomed in the y-axis. Here the unit of $\Delta$ is Planck mass $M_p$. \label{222} }
\end{figure}

In the Fig.~\ref{222} we study the bifurcation probability numerically for large field inflation. The pink region is the constraint by stuck or unphysical ending of our numerical calculation (in which case the field runs too far away in the isocurvature direction, beyond the size of the grid)
, which is beyond the scope of our interest.

Note that in the 3, 5 and 10 dimensional cases in Fig.~\ref{222}, the stuck region (pink) does not fully cover the parameter space from above. To understand this, it is helpful to take a closer look at how the trajectory can get stuck in a highly random potential. To get stuck, the inflaton typically first falls into a potential well from one side, and then cannot climb out from the other side. Note that the inflaton first obtains some kinetic energy when it falls into a potential well. After that, there are two cases when the inflaton get stuck and cannot get out:
\begin{enumerate}
\item The potential well at ``the other side'' is higher than the potential well at ``one side'' plus the inflaton kinetic energy (calculated before falling into the potential well). In this case, there is no way for the inflaton to classically climb up (and the quantum effect is small) from energy conservation.
\item The potential well at ``the other side'' is higher than the inflaton kinetic energy; but not as large as the kinetic energy (calculated before falling into the potential well) plus the energy obtained from falling into the potential well from the ``one side''. In this case, whether or not the inflaton gets stuck depends on the feature distance $\Delta$. This is because, when $\Delta$ is small, the fall-climb happens very fast and the fraction force from Hubble parameter does not play a significant role. In this case, the inflaton can climb out following energy conservation. However, when $\Delta$ is large enough, the additional inflaton kinetic energy which is obtained from falling into the potential well gets diluted from Hubble expansion. As a result, the inflaton cannot climb out if the potential well is higher than the inflaton kinetic energy.
\end{enumerate}

Note that case 2 is less demanding to be triggered. Thus the stuck probability shrinks when the feature distance $\Delta$ is small.

\begin{figure}[htbp]
  \centering
  \includegraphics[width=0.8\textwidth]{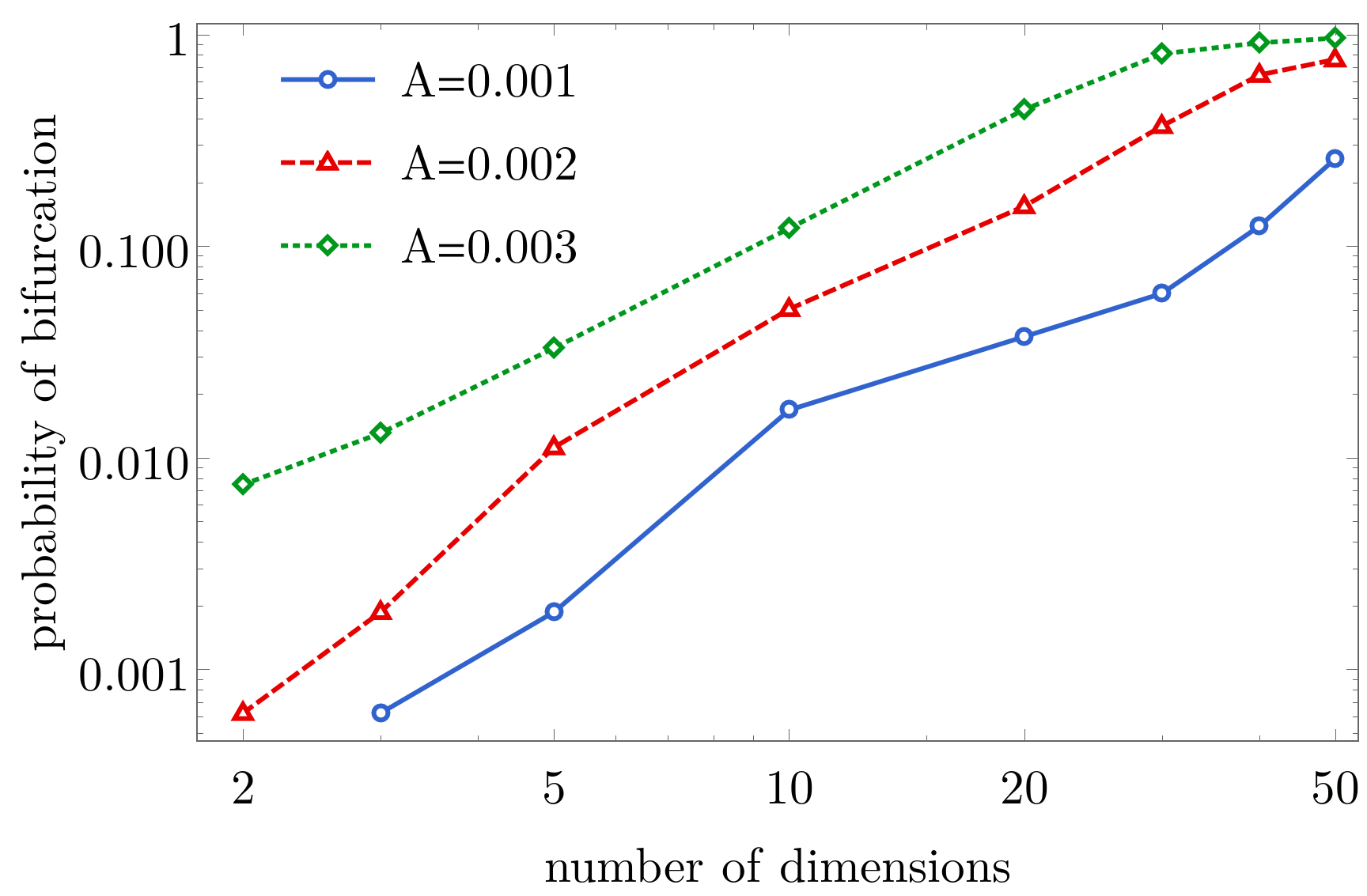}
  \caption{\label{fig:dim-large-log} The bifurcation probability as a function of number of field-space dimensions, for large field inflation.}
\end{figure}

To better observe the scaling behavior as a function of number of fields, we fix $\Delta$ at 0.1$M_p$ and consider a few examples of amplitudes. The result is shown in Fig.~\ref{fig:dim-large-log}. One can clearly observe that the increment of bifurcation probability is faster than linear as field-space dimension increases (except that when the bifurcation probability is close to 1, the rapid increasing behavior certainly stops). This behavior is natural to expect. Because linear increment simply reflects the fact that there are more directions to have bifurcation events. On the other hand, bifurcation may happen in a (time dependent) linear combination of different directions, which further increase the bifurcation probability. As one can read from the figure, the increment behavior is approximately quadratic.

\begin{figure}[htbp]
  \centering
  \includegraphics[width=0.8\textwidth]{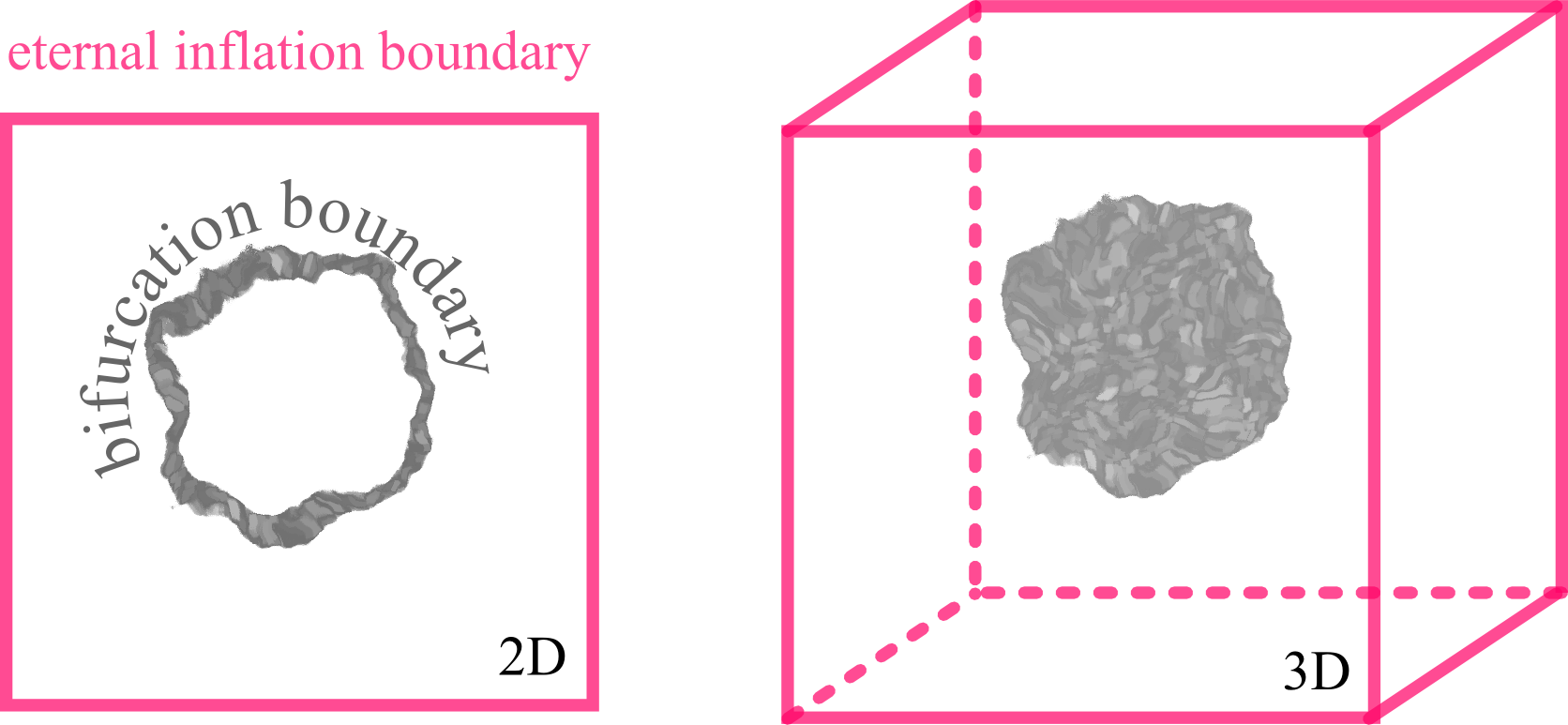}
  \caption{\label{fig:high-d-volume} In high field-space dimensions, the parameter space of non-bifurcating region takes an exponentially small volume. Here the pink boundary corresponds to the parameter space of non-eternal inflation. The gray ball-like shape corresponds to the non-bifurcation region. Note that this figure is for illustration of the scaling rule, which is not calculated from real data.}
\end{figure}

Finally, let us estimate the fraction of non-eternally-inflating parameter space which bifurcates. To actually calculate such a fraction of parameter space requires exponentially computational complexity. Because we need to fill the $n$-dimensional parameter space $A_i$ ($i=0, 1, \ldots, n$) with numbers, where $A_i$ can now take different values for different $i$. However, one can nevertheless estimate the parameter space by noting that the non-bifurcating region is a ball-like sub-volume of the non-eternally-inflating parameter space. Thus the volume ratio is exponentially small. The situation is illustrated in Fig.~\ref{fig:high-d-volume}. In general, for $N$ dimensions, the volume fraction $f$ scales as
\begin{align}
  f = \left(\frac{A_b}{A_e}\right)^{N}~,
\end{align}
where $A_b$ is the amplitude boundary for bifurcation (where more than $10\%$ of the trajectories bifurcate), and $A_e$ is the amplitude boundary for eternal inflation.

Note that with multiple dimensions, eternal inflation become unlikely because with more directions to go, the field is less likely to get stuck. Thus the non-eternal parameter space tends to be larger than the exponential estimate. However, with multiple dimensions, the probability of bifurcation is increasing because every direction, and their linear combinations, may bifurcate. Thus the non-bifurcating parameter space tends to be smaller than the exponential estimate. As a result, it is safe to estimate that the non-bifurcating parameter space is an exponentially small subset of the non-eternally inflating parameter space, as a function of $N$.

\begin{figure}[htbp]
  \centering
  \includegraphics[width=0.8\textwidth]{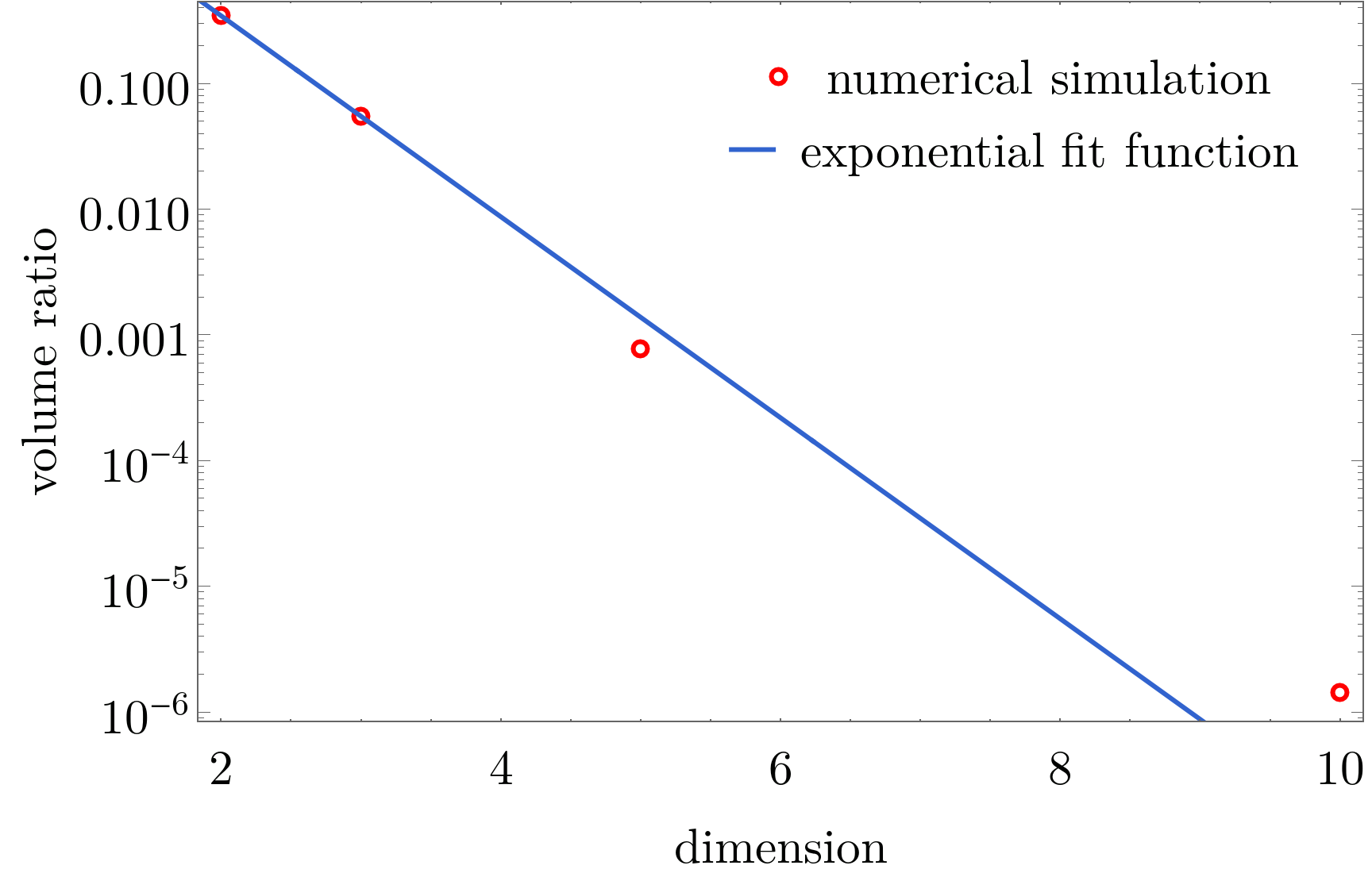}
  \caption{\label{fig:ratio} An estimate of the non-bifurcating to non-eternal inflating volume ratio for different dimensions for large field inflation.}
\end{figure}

To put in real data, the estimate of bifurcation probability is plotted in Fig.~\ref{fig:ratio}. For example, in 10 dimensions, only about $10^{-6}$ of the non-eternal inflation parameter space does not suffer from bifurcations.

\subsection{Small field inflation}

For small field inflation, we consider a brane inflation \cite{BraneInflation} potential
\begin{equation}
V=V_0(1-\frac{\mu^4}{\phi_1^4})~,
\end{equation}
plus a small random potential. The parameters are taken according to \cite{HBraneInflation}, with $\mu=0.01 M_p$.

\begin{figure}[htbp]
  \centering
  \includegraphics[width=0.8\textwidth]{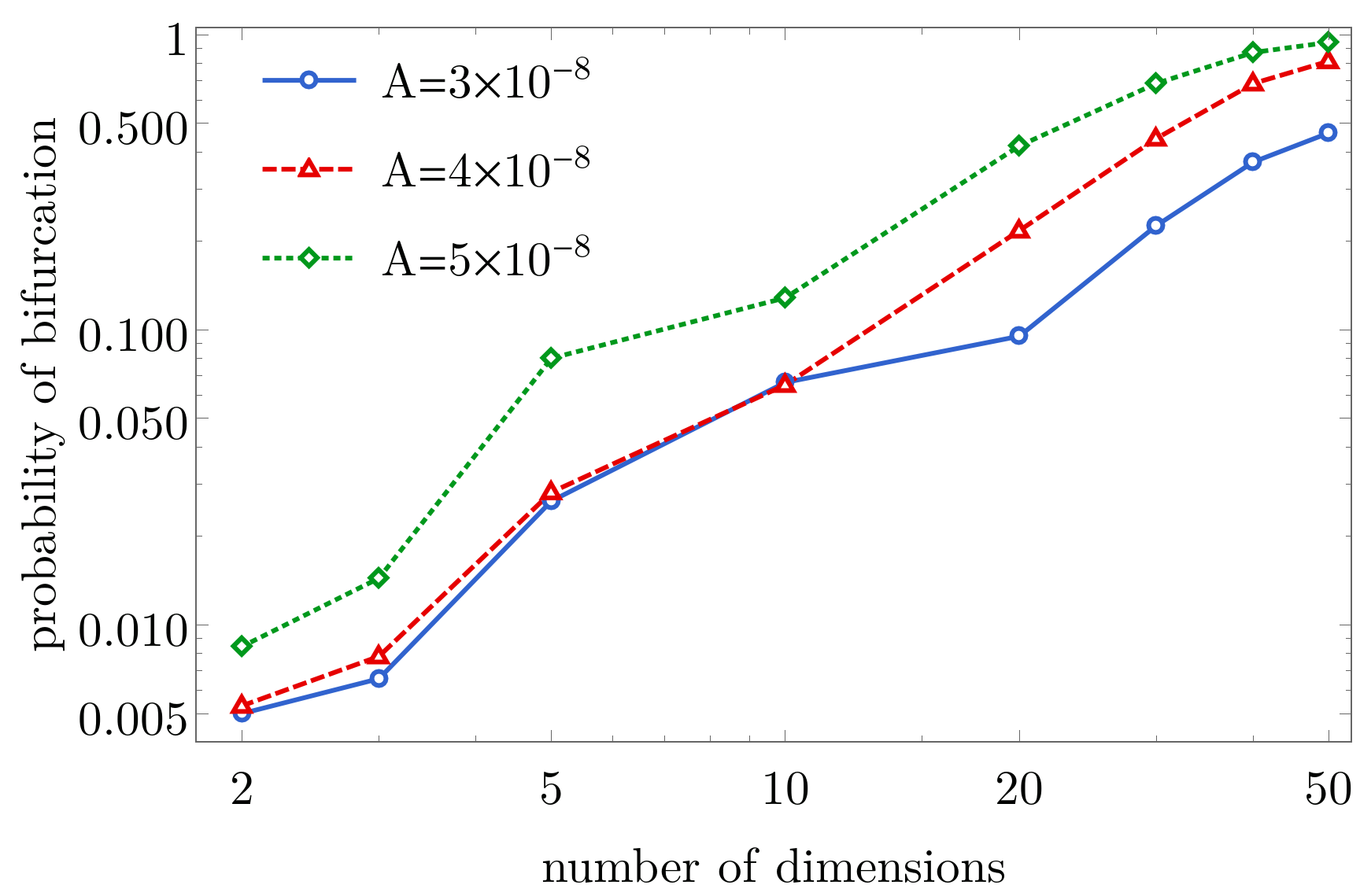}
  \caption{\label{fig:dim-small-log} The bifurcation probability as a function of number of field-space dimensions, for small field inflation. }
\end{figure}

\begin{figure}[htbp]
  \centering
  \includegraphics[width=0.8\textwidth]{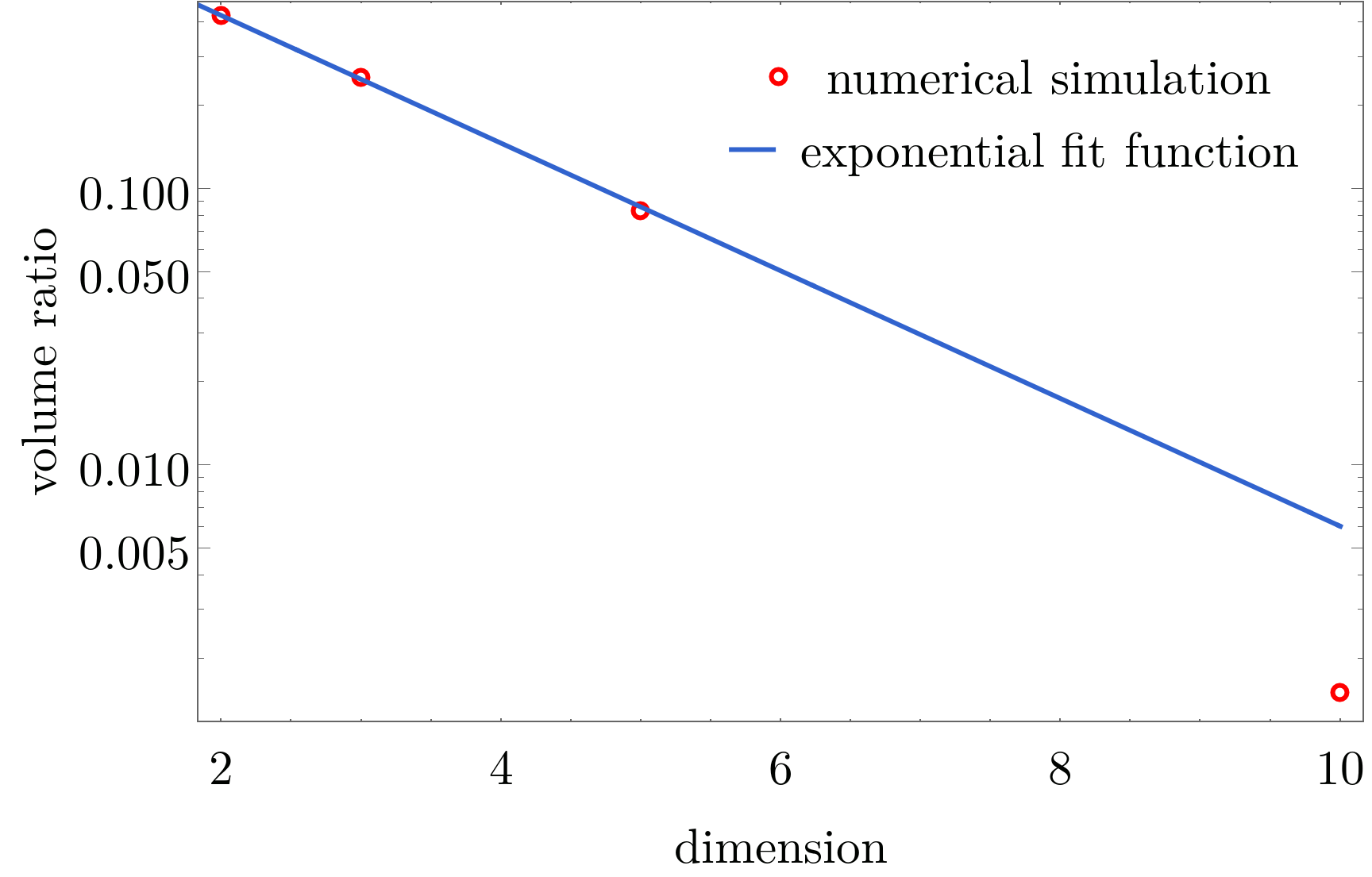}
  \caption{\label{fig:ratio-small} An estimate of the non-bifurcating to non-eternal inflating volume ratio for different dimensions for small field inflation.}
\end{figure}

The bifurcation probability as a function of number of field-space dimensions is plotted in Fig.~\ref{fig:dim-small-log}, where $\Delta=0.0005 M_p$ is chosen. Similar to the large field case, a faster-than-linear growth is spotted. On the other hand, bifurcation requires much smaller $A$ in small field inflation than large field inflation, because the field is rolling more slowly, and thus more sensitive to features in the potential.  The ratio of non-bifurcating to non-eternal inflationary parameter spaces is estimated in Fig.~\ref{fig:ratio-small}. It is noted that in small field inflation, the reduce of parameter space is not that dramatic as the case of large field inflation. Nevertheless, only about $10^{-3}$ of the total non-eternal inflationary parameter space is free from bifurcations.

\section{Conclusion and outlook} \label{sec:conclusion-outlook}
To conclude, after proposing a new way to approximately construct a random multi-field inflationary potential, we extend our previous study of bifurcation phenomena into the case of higher field space dimensions. The situation turns out to be not a qualitative correction, but instead a remarkable quantitative change. There are two sources of enhancements of the bifurcation probability:
\begin{itemize}
\item Enhancement of bifurcation probability as a function of field-space dimensions, when fixing other potential parameters. As we have shown, when we add field directions, the bifurcation probability increases. The speed of increment is faster than linear in both the large field and the small field examples that we have studied. This result is natural to expect. This is because a linear behavior should be the lower bound of bifurcation probability as a function of field dimension, simply considering that there are more directions to move into. On the other hand, bifurcation could either happen on a linear combination of those field dimensions, which makes the behavior faster than linear.
\item A ball-like sub-volume in high dimensions has exponentially tiny space fraction compared to the whole parameter space. We do not know the exact volume ratio between the non-bifurcating sub-volume and the full parameter space. Because one has to simulate a exponentially high number of potentials with different parameters to really account for this ratio. However, the ratio should scale exponentially as a number of dimensions. Thus only an exponentially small fraction of parameter space is left without bifurcations, which has a chance to agree with any observations.
\end{itemize}

In the present paper, we have not applied constraints from observation, like features in the primordial power spectrum, and non-Gaussianities, to limit the shape of the inflationary potential. Other than technical simplicity, we have the following considerations for not mix-in the latest observations:
\begin{itemize}
\item The current observational constraints on highly scale dependent features are not yet very strong. Especially, the current data-fitting typically focuses on the encounter of a single feature instead of averaging a large number of features.
\item Most of the experiments, especially the most precise ones, observes only the first 10 e-folds of inflation, among the 50-60 e-folds along the ``observable'' inflationary trajectory. On the other hand, our bifurcation constraint covers a much greater number of e-folds because otherwise primordial black holes could have been formed from the density fluctuation of the bifurcated trajectory (but those primordial black holes cannot be predicted if one followed only one classical trajectory), which contradicts observations.
\item In numerical studies of observational features of random potentials (such as non-Gaussianities), one typically first generate a random potential, and test that the potential does not lead to eternal inflation. Then a statistical study of the observational features is performed (see, for example, \cite{Dias:2012nf}). Our result show that, an exponentially large portion of parameter space should be cut off well before hitting the eternal inflation constraint.
\end{itemize}
On the other hand, it is indeed interesting to combine all observational constraints and we hope to do so in future work.

\section*{Acknowledgments}
YW was supported by a Starting Grant of the European Research Council (ERC STG grant 279617), and the Stephen Hawking Advanced Fellowship. SZ is supported by the the Hong Kong PhD Fellowship Scheme (HKPFS) issued by the Research Grants Council (RGC) of Hong Kong. We would also like to thank the Institute for Advanced Study, Hong Kong University of Science and Technology, where part of this work is finished.


\begin{thebibliography}{999}

\bibitem{Berera:1996nv}
  A.~Berera,
  Phys.\ Rev.\ D {\bf 54}, 2519 (1996)
  [hep-th/9601134].

\bibitem{Huang:2008jr}
  Q.~G.~Huang and S.-H.~H.~Tye,
  Int.\ J.\ Mod.\ Phys.\ A {\bf 24}, 1925 (2009)
  [arXiv:0803.0663 [hep-th]].

\bibitem{Tye:2009ff}
  S.-H.~H.~Tye and J.~Xu,
  Phys.\ Lett.\ B {\bf 683}, 326 (2010)
  [arXiv:0910.0849 [hep-th]].

\bibitem{Battefeld:2011yj}
  D.~Battefeld, T.~Battefeld, C.~Byrnes and D.~Langlois,
  JCAP {\bf 1108}, 025 (2011)
  [arXiv:1106.1891 [astro-ph.CO]].

\bibitem{Dias:2012nf}
  M.~Dias, J.~Frazer and A.~R.~Liddle,
  JCAP {\bf 1206}, 020 (2012)
  [Erratum-ibid.\  {\bf 1303}, E01 (2013)]
  [arXiv:1203.3792 [astro-ph.CO]].

\bibitem{McAllister:2012am}
  L.~McAllister, S.~Renaux-Petel and G.~Xu,
  JCAP {\bf 1210}, 046 (2012)
  [arXiv:1207.0317 [astro-ph.CO]].

\bibitem{Gao:2014uha}
  X.~Gao, T.~Li and P.~Shukla,
  JCAP {\bf 1410}, no. 10, 048 (2014)
  [arXiv:1406.0341 [hep-th]].

\bibitem{Green:2014xqa}
  D.~Green,
  arXiv:1409.6698 [hep-th].

\bibitem{Duplessis:2012nb}
  F.~Duplessis, Y.~Wang and R.~Brandenberger,
  JCAP {\bf 1204}, 012 (2012)
  [arXiv:1201.0029 [hep-th]].

\bibitem{Tegmark:2004qd}
  M.~Tegmark,
  JCAP {\bf 0504}, 001 (2005)
  [astro-ph/0410281].

\bibitem{Frazer:2011br}
  J.~Frazer and A.~R.~Liddle,
  JCAP {\bf 1202}, 039 (2012)
  [arXiv:1111.6646 [astro-ph.CO]].

\bibitem{Marsh:2013qca}
  M.~C.~D.~Marsh, L.~McAllister, E.~Pajer and T.~Wrase,
  JCAP {\bf 1311}, 040 (2013)
  [arXiv:1307.3559 [hep-th]].

\bibitem{Bachlechner:2014rqa}
  T.~C.~Bachlechner,
  JHEP {\bf 1404}, 054 (2014)
  [arXiv:1401.6187 [hep-th]].


\bibitem{Battefeld:2014qoa}
  T.~Battefeld and C.~Modi,
  arXiv:1409.5135 [hep-th].

\bibitem{Li:2009sp}
  M.~Li and Y.~Wang,
  JCAP {\bf 0907}, 033 (2009)
  [arXiv:0903.2123 [hep-th]].

\bibitem{Wang:2013vxa}
  Y.~Wang,
  JCAP {\bf 1310}, 006 (2013)
  [arXiv:1304.0599 [astro-ph.CO]].

\bibitem{Afshordi:2010wn}
  N.~Afshordi, A.~Slosar and Y.~Wang,
  JCAP {\bf 1101}, 019 (2011)
  [arXiv:1006.5021 [astro-ph.CO]].

\bibitem{Jazayeri:2014nya}
  S.~Jazayeri, Y.~Akrami, H.~Firouzjahi, A.~R.~Solomon and Y.~Wang,
  JCAP {\bf 1411}, 044 (2014)
  [arXiv:1408.3057 [astro-ph.CO]].

\bibitem{Li:2009me}
  S.~Li, Y.~Liu and Y.~S.~Piao,
  Phys.\ Rev.\ D {\bf 80}, 123535 (2009)
  [arXiv:0906.3608 [hep-th]].

\bibitem{Wang:2010rs}
  Y.~Wang,
  Journal of Cosmology, {\bf 4}, 744-759 (2010).
  [arXiv:1001.0008 [hep-th]].

\bibitem{Long:2014dta}
  C.~Long, L.~McAllister and P.~McGuirk,
  Phys.\ Rev.\ D {\bf 90}, 023501 (2014)
  [arXiv:1404.7852 [hep-th]].

\bibitem{Chen:2009we}
  X.~Chen and Y.~Wang,
  Phys.\ Rev.\ D {\bf 81}, 063511 (2010)
  [arXiv:0909.0496 [astro-ph.CO]].

\bibitem{Chen:2009zp}
  X.~Chen and Y.~Wang,
  JCAP {\bf 1004}, 027 (2010)
  [arXiv:0911.3380 [hep-th]].

\bibitem{Starobinsky:1986fx}
  A.~A.~Starobinsky,
  Lect.\ Notes Phys.\  {\bf 246}, 107 (1986).

\bibitem{BraneInflation}
S. H. H. Tye, Lect.\ Notes Phys.\ {\bf 737}, 949 (2008),
arXiv:hep-th/0610221.

\bibitem{HBraneInflation}
Yin-Zhe Ma, Qing-Guo Huang, Xin Zhang
[arXiv:1303.6244[astro-ph.CO]]





\end{thebibliography}
\end{document}